\documentclass[journal,transmag]{IEEEtran}

\usepackage{graphicx}
\usepackage{amsmath}
\usepackage{mathrsfs}
\usepackage{amsfonts}
\usepackage{amssymb}

\DeclareFontFamily{OT1}{phv}{}
\DeclareFontShape{OT1}{phv}{m}{n}{<-> s * [0.9] phvr7t}{}
\DeclareFontShape{OT1}{phv}{m}{it}{<-> s * [0.9] phvro7t}{}

\begin{document}

\title{Using Modularity Metrics to assist Move Method
Refactoring of Large Systems}

\author{\IEEEauthorblockN{Christian
Napoli\IEEEauthorrefmark{1,*},
Giuseppe Pappalardo \IEEEauthorrefmark{1}, and
Emiliano Tramontana\IEEEauthorrefmark{1}}
\IEEEauthorblockA{\IEEEauthorrefmark{1}Dpt. of Mathematics and
Computer Science, University of Catania, Italy}% <-this % stops
an unwanted space
\thanks{*Email: napoli@dmi.unict.it.}
\thanks{PUBLISHED ON: \bf IEEE Transactions on Neural Networks
and Learning Systems, vol. 23, N. 11, pp. 1805-1815, 2012}}

% The paper headers
\markboth{Using Modularity Metrics to assist Move Method
Refactoring of Large Systems -- PREPRINT}%
{Shell \MakeLowercase{\textit{et al.}}: Bare Demo of
IEEEtran.cls for Journals}

 \begin{titlepage}
 \begin{center}
 {\Large \sc PREPRINT VERSION\\}
  \vspace{15mm}
{\Huge Using Modularity Metrics to assist Move Method
Refactoring of Large Systems\\}
 \vspace{10mm}
 {\Large C. Napoli, G. Pappalardo, and E. Tramontana\\}
 \vspace{15mm}
{\Large \sc PUBLISHED ON: \bf 7th International Conference on Complex, Intelligent, and Software Intensive Systems (CISIS), pp. 529-534, 2013\\}
 \end{center}
 \vspace{10mm}
 {\Large \sc BIBITEX: \\}
 \vspace{5mm}
 
@inproceedings\{Napoli20132013, \\
author=\{Napoli, C. and Pappalardo, G. and Tramontana E.\}, \\
booktitle=\{2013 7th International Conference on Complex, Intelligent, and Software Intensive Systems (CISIS)\}, \\
title=\{Using Modularity Metrics to assist Move Method
Refactoring of Large Systems\}, \\
year=\{2013\}, \\
pages=\{529-534\}, \\
publisher=\{IEEE\}, \\
doi=\{10.1109/CISIS.2013.96\}, \\
\}
 \vspace{35mm}
 \begin{center}
Published version copyright \copyright~2013 IEEE \\
\vspace{5mm}
UPLOADED UNDER SELF-ARCHIVING POLICIES\\
NO COPYRIGHT INFRINGEMENT INTENDED \\
 \end{center}
\end{titlepage}

\title{Using Modularity Metrics to assist Move Method
Refactoring of Large Systems}

% Refactoring Large Software Systems by means of Modularity
%  Metrics}

%Suggesting Refactoring on the basis of Modularity Metrics
%  computed for Large Software Systems}

\author{\IEEEauthorblockN{Christian Napoli, Giuseppe Pappalardo,    Emiliano Tramontana}
  \IEEEauthorblockA{Dipartimento di Matematica e Informatica\\
    University of Catania, Italy\\
    Email: \{napoli, pappalardo, tramontana\}@dmi.unict.it }}

\maketitle

\begin{abstract}
  For large software systems, refactoring activities can be a
  challenging task, since for keeping component complexity under
  control the overall architecture as well as many details of each
  component have to be considered.  Product metrics are therefore
  often used to quantify several parameters related to the modularity
  of a software system.

  This paper devises an approach for automatically suggesting
  refactoring opportunities on large software systems.  We show that
  by assessing metrics for all components, move methods refactoring
  can be suggested in such a way to improve modularity of several
  components at once, without hindering any other.
  However, computing metrics for large software systems, comprising
  thousands of classes or more, can be a time consuming task when
  performed on a single CPU.  For this, we propose a solution that
  computes metrics by resorting to GPU, hence greatly shortening
  computation time.

  Thanks to our approach precise knowledge on several properties of
  the system can be continuosly gathered while the system evolves,
  hence assisting developers to quickly assess several solutions for
  reducing modularity issues.
\end{abstract}


%% in introduzione inserire che il principio di move method puo'
%% migliorare diverse metriche, in modi diversi

\section{Introduction}

For assessing and enhancing the modularity of existing object-oriented
systems, several techniques have been proposed~\cite{Gamma94,
  Fowler99, PacoSac06}.  Refactoring techniques, e.g.\ extract
snippets of code from a method, move a method to a different class,
etc., are a fundamental support for improving the cohesion of a class,
reducing coupling between classes, shortening long methods/classes,
etc.~\cite{Fowler99, Kerievsky05}.
Accordingly, \emph{code smells}, i.e.\ characteristics indicating that
\emph{``the code has to be changed''}~\cite{Fowler99}, have been
proposed to guide refactoring activities.
Moreover, several well-known \emph{metrics} have been proposed to
assess the characteristics of software systems and to guide
refactoring~\cite{HenryK81, ChidamberK94,
% Lanza06, 
  SarkarKR08, TsantalisC09}.
Indeed, using several metrics at once, for a large software system,
with each metric assessing a different facet, can give the developer a
better understanding of modularity and quality characteristics.

In our view, refactoring should help to meet more than one modularity
need and assessing multiple metrics at once would be essential when
trying to balance several \emph{``forces''} acting into a software
system.

For \emph{crosscutting concerns}, i.e.\ concerns that have been
implemented by spreading the correspondent code across several
modules, refactoring can make good use of \emph{aspect-oriented
  programming} (AOP)~\cite{Kiczales97}.  
% When using AOP, snippets of
% code can be moved out of classes and gathered as separate aspects that
% intervene into selected operations performed by classes. 
Refactoring to aspects has been proposed in order to extract method
calls, conditional statements, etc.~\cite{BinkleyCHRT06, Marin07} or
in order to have separation between
%well-known design patterns implemented without mixing
domain and pattern-related code~\cite{GiuntaPT12} or non-functional
requirements~\cite{GiuntaMPT12aug, Kaqudai}.
%An analysis of the
%usage of aspect technology can e.g.\ be found in~\cite{apel2010aspectj}.  
Accordingly, some metrics have been proposed to evaluate how concerns
spread over modules (i.e.\ classes or aspects)~\cite{MetricSac05,
  Kulesza06}.
%  Such metrics are based on
% data collected from modules and concerns.
 
This paper aims at tackling two issues related to the suggestion of
move method refactoring opportunities.  Firstly, 
%having an approach whereby 
%proposing a combination of metrics 
automatically finding
% indicates a
refactoring suggestions that improve several components at once, which
becomes possible by computing several metrics.
E.g.\
cohesion for a pair of components is improved at once by carefully
selecting a single move method refactoring, without hindering other
characteristics of the same pair and without negatively affecting
other components.  How a change  affects components
 is unclear when having to cope with a
large number of them.
%  since it  means keeping in mind all
% the details of each  affected part.
%
% Hence, by computing several product metrics simultaneously, we are
% able to suggest refactoring opportunities, while taking into account
% and improving different facets of modularity.
%
Automatic suggestion of refactoring opportunities 
is of great importance for large systems, since the
unassisted developer, having to reason on thousands of classes, could
miss the proper refactoring. Moreover, the number of ways in which
\emph{perfective} changes can be introduced is dramatically increased
in such large systems.  Manual exploration of such changes would be
  cumbersome or time consuming.

Secondly, computing metrics should take a tiny amount of time, when a
software system consists of a large number of methods, attributes, and
classes, i.e.\ in the order of several thousands, to be considered a
useful indication to the developer while she explores refactoring
opportunities.  For this, we have devised a parallel algorithm that
runs on a GPU to compute time consuming product metrics and we have
greatly reduced, even by a factor of 50, the typical CPU computing
time needed.
A GPU provides hundreds of computing cores, whereas a CPU provides a
few (typically 8). To effectively employ hundreds of cores, our
solution let threads run without synchronisation as much as possible.

The remaining part of the paper is organised as follows.
Section~\ref{metrics} introduces relevant metrics assessing the
modularity of software systems.  Section~\ref{approach} proposes how
to combine metrics for suggesting the most appropriate refactoring and
how computationally costly this can be.  Section~\ref{computing}
describes our solution for collecting data on systems, as well as for
computing metrics on GPUs.  Section~\ref{caseStudies} reports the
results of our experiments on some real-world systems.  Finally,
concluding remarks are drawn in Section~\ref{conclusions}.

\section{Modularity Metrics}
\label{metrics}

%Many metrics have been proposed for assessing modularity.  
Among the most useful modularity metrics, in our approach we compute
the following ones.

\subsection{Structural metrics}
Henry and Kafura proposed \emph{Fan-in} and \emph{Fan-out} for
procedural programming~\cite{HenryK81}, however these are also
used for object-oriented systems.
\emph{Fan-in} counts for a method $m$ the number of methods calling
$m$.  A method having a high value of fan-in encloses a fragment of
code that has been reused many times within the system itself.  For
this, a change on it could trigger many changes on parts of the system
relying on it, i.e.\ suffer of the \emph{ripple effect}.  A method
with high fan-in should be carefully tested to ensure that calling
methods rely on its correct behaviour.
High fan-in values can indicate crosscutting concerns, and accordingly
method calls could be extracted and implemented as
aspects~\cite{Marin07}.

\emph{Fan-out} counts for a method $m$ the number of methods called by
$m$. A method with a high value of fan-out can be considered complex
and having too many responsibilities. Hence, it is difficult to reuse,
because it depends on a large number of methods.
A method exhibiting a large fan-out, having a large amount of
dependencies, should be split into several methods, each constituting
a lower-level abstraction for the caller~\cite{Gamma94, Fowler99, MessinaPT07}.

For a pair of entities, \emph{Jaccard similarity coefficient} measures
the ratio between the intersection cardinality of the sets of
properties for each entity, and the union cardinality of the two
sets~\cite{TsantalisC09}.  For a method, its properties can be defined
as called methods and accessed attributes.  Hence, the similarity for
a pair of methods gives a measure of how close they should be, i.e.\
whether it is appropriate to have such methods on the same class or
not~\cite{SimonSL01, TsantalisC09}.
%
%Therefore, such a metric suggests move methods opportunities, i.e.\
Methods pairs with high similarity are better handled on the same
class. This is because, in general, a high degree of intra-class
interactions suggests high cohesion for the class, whereas for a pair
of methods belonging to different classes, the high degree of coupling
hinders modularity.

\subsection{Object-oriented metrics}
Chidamber and Kemerer have introduced a widely used suite of metrics
that provides several parameters for assessing the modularity of
object-oriented systems~\cite{ChidamberK94}.  The well-known suite
comprises \emph{WMC}, \emph{NOC}, \emph{DIT}, \emph{CBO}, \emph{RFC}
and \emph{LCOM}.
In this paper we focus on \emph{CBO} and \emph{LCOM}.

\emph{CBO} measures \emph{coupling between objects} by counting for a
class $C$ the number of other classes used by $C$, hence method calls
or instantiations of classes, etc.\ performed by $C$.
Similarly to \emph{fan-out}, high values of \emph{CBO} indicate
complexity of classes, i.e.\ too many dependencies, and as such it
would be better to refactor.

\emph{LCOM} measures \emph{lack of cohesion on methods} of a class $C$
and has been originally defined as the number of method pairs sharing
no attributes minus the number of method pairs that share at least one
attribute.
Later, it has been proposed to measure LCOM as 1 minus the average,
among methods of $C$, of the ratio beween the number of attributes of
$C$ used by each method and the total number of attributes of
$C$~\cite{BriandDW98}.
% inserire riferimento
% Forse e' Henderson-Sellers, vedi Briand, Daly, Wust. Unified
% Framework for cohesion ... 1998
% $$ LCOM(C) = 1- average |a: m uses a| /  |a : a is in C| $$
%
For a class with a high value of \emph{LCOM}, its methods have been
incidentally put together, hence some methods should be moved to
another class, or some attributes should be moved into this class.

% \subsection{Concern-oriented metrics}

% \emph{Coupling between Concerns (CBC)} has been defined as the number of
% calls by a group of classes, which have been characterised with the
% same concern, to different methods belonging to a class (one or more
% classes) of a different concern, i.e.\ group~\cite{MetricSac05}.
% This metrics aims at revealing how concerns are related with each other.
% %
% Grouping classes belonging to the same concern can be considered a
% decision that depends on the point of view of the developer performing
% an analysis. However, for large software systems, this decision can be
% a burden. In the following, we have assumed that groups of classes
% have been defined according to the correspondent Java packages of
% classes.

% \emph{Fragm} for a class has been defined as the concern heterogeneity
% of the class. Suppose a class has snippets of code belonging to
% different categories, then a class has a value of fragmentation that
% grows with the number of categories~\cite{MetricSac05}.  Fragm has
% been quantified as a sort of distance
% %scalar product
% %et: vedi scalar product
% between the vectors representing the different concerns for a class
% (see~\cite{MetricSac05} for the exact formulation).
% %et: vedi formula
% We have considered the categories for snippets of code suggested by
% the paper proposing the metrics, which can be automatically attributed
% to the code.

\section{Approach for refactoring}
\label{approach}

\subsection{How refactoring is suggested}
By evaluating the above metrics simultaneously, we advocate that the
modularity of an object-oriented system can be improved by means of
the following indications.
% and activities.

%uso: lcom, similarity, fan-out
When the similarity between a pair of methods ($m_1$, $m_2$) is
high\footnote{Let us say that a threshold is chosen for the similarity
  value as e.g.\ the average among all similarity values for the whole
  system.} and the methods happen to be on different classes, we
check how LCOM values for classes varies when moving one method from its origin
class to the class of the other method on the pair, or to other
classes holding one of the called methods.
Then, we suggest to move a method, i.e.\ $m_1$ or $m_2$, from its
origin class to another class when we find that 
LCOM of both the origin and destination classes will be improved.

%uso lcom, fan-out
When LCOM for a class is high, hence methods happen to incidentally be
on such a class, then an improvement is given by taking out one (or
more) method from the class and find another class for it.
Tentative destination classes for a method will be the ones holding
methods called (or attribute accessed) by the method to be
moved. Methods with the highest values of fan-out will be selected as
methods to be moved out from the origin class.
The class suggested as a destination for a method will be the one
whose LCOM will be lowered, while also lowering LCOM of the origin
class.  Note that, a desired side-effect is that CBO of the origin
class could be lowered by such a suggested refactoring.

%uso: cbo, lcom, fan-out, similarity
When CBO for a class is high, we check possible methods relocations,
starting with the ones having the highest fan-in and fan-out, to
another class.  Candidate destination classes are all the classes used
by the method to be moved.
CBO will be computed for candidate destination classes, for each
possible relocation.
The proposed refactoring selects as a destination for a method, the
class whose CBO will be not higher than its original value, while of
course the CBO of the origin class will be lowered.
The granularity of the desired improvement can be decided by setting a
threshold for the desired CBO (e.g.\ the threshold could be set as the
average value of CBO for all classes).

% A class having a high value of fragm can be improved by moving out a
% method from the class. The candidate destination class of the method
% is the one having already the same concern of the method to be
% moved. Candidate destination classes are classes of the same group of
% the origin class, and the group is known thanks to the definition of
% CBC.
% %
% The proposed destination class for a method is the one whose value
% of fragm will be not higher than that before moving a method in.

For large software systems, computing metrics  for all the
tentative methods moves is time consuming unless appropriate high
performance resources are employed.

\subsection{How computing metrics scales}
\label{scales}

This section estimates the number of metric values computed when using
the above approach to suggest refactoring opportunities.
Firstly, fan-in and fan-out will be computed for every method.
Suppose $m$ is the number of methods, then the number of values for
each of the two metrics is the number of available methods $m$.

Note that, as far as complexity is concerned, for computing fan-in for
a method $m_p$, all other $m-1$ methods will have to be scanned, and
each invoked method will have to be compared with method $m_p$. Let us
suppose that maximum $k_m$ methods are called by each method, then the
worst case is that $k_m \cdot (m-1)$ comparisons have to be performed.

Secondly, similarity will be computed on all possible combinations of
methods pairs.  From combinatorics, given $m$ methods, then the number
of all pairs, without considering the ordering, is given by
$$\textrm{\#Similarity values} = {m \choose 2} = \frac{m \cdot (m-1)}{2}$$
From an analysis likewise the one above would come out that for a method
pair the number of comparisons to be performed are $(k_m+k_a)^2$ in
the worst case, where $k_a$ is the maximum number of attributes
accessed by each method.
Computing similarity \emph{sequentially} for thousands of methods can
be time consuming, since the number of pairs quickly grows with the
square of the number of methods.

Thirdly, LCOM and CBO will be computed for each class, as well
as on each candidate class that could receive or give up a method.
For $c$ classes, and $m$ methods, the worst case is to assess LCOM and
CBO after having moved into each class every method, as well as having
moved out a method from each class.
The number of LCOM values is 
$$\textrm{\#LCOM values} = c + c \cdot m = c \cdot (m+1)$$
For thousands of methods and classes the said number of values grows
as $c \cdot m$, hence computing it \emph{sequentially} can be time
consuming.
The number of CBO values is the same as the above number of LCOM
values, in the worst case.

% Finally, for CBC the number of values to compute depend on the
% number of concerns into which we can group classes. Suppose that $c$
% classes have been grouped into $g$ concerns, the number of pairs of
% concerns is given by combinatorics as $g \cdot (g-1)/2$.
% Finally,  fragm will have to be computed for each class as well as for
% all the origin and candidate destination classes for assessing a
% method move. The number of fragm values will be as for LCOM values. 

Therefore, for the whole set of metrics presented above, we have that
the number of values to  compute is
$$ \# \textrm{values} = 2 \cdot m + \frac{m \cdot (m-1)}{2} + 2 \cdot
c \cdot (m+1) $$
%+ \frac{g \cdot (g-1)}{2} $$
%
The second amount of the sum grows quicker than the others (while the
third grows quicker than the first) for increasing values of $m$. Note
also that, generally, $m$ grows quicker than $c$, for practical
software systems.

Let us consider a software system having $1,200$
methods and $231$ classes, this is the case for JUnit, which is
considered  a small software system.  Then, the number of values to
compute are $1.2$ \emph{millions}.
Section~\ref{caseStudies} shows the number of classes and methods, as
well as measured computing times, for several real-world software
systems.

\section{Analysing Software Systems}
%Computing Metrics}
\label{computing}
% \subsection{Organising application data}

\subsection{Overview}

\begin{figure}
  \begin{center}
    \includegraphics[scale=0.46]{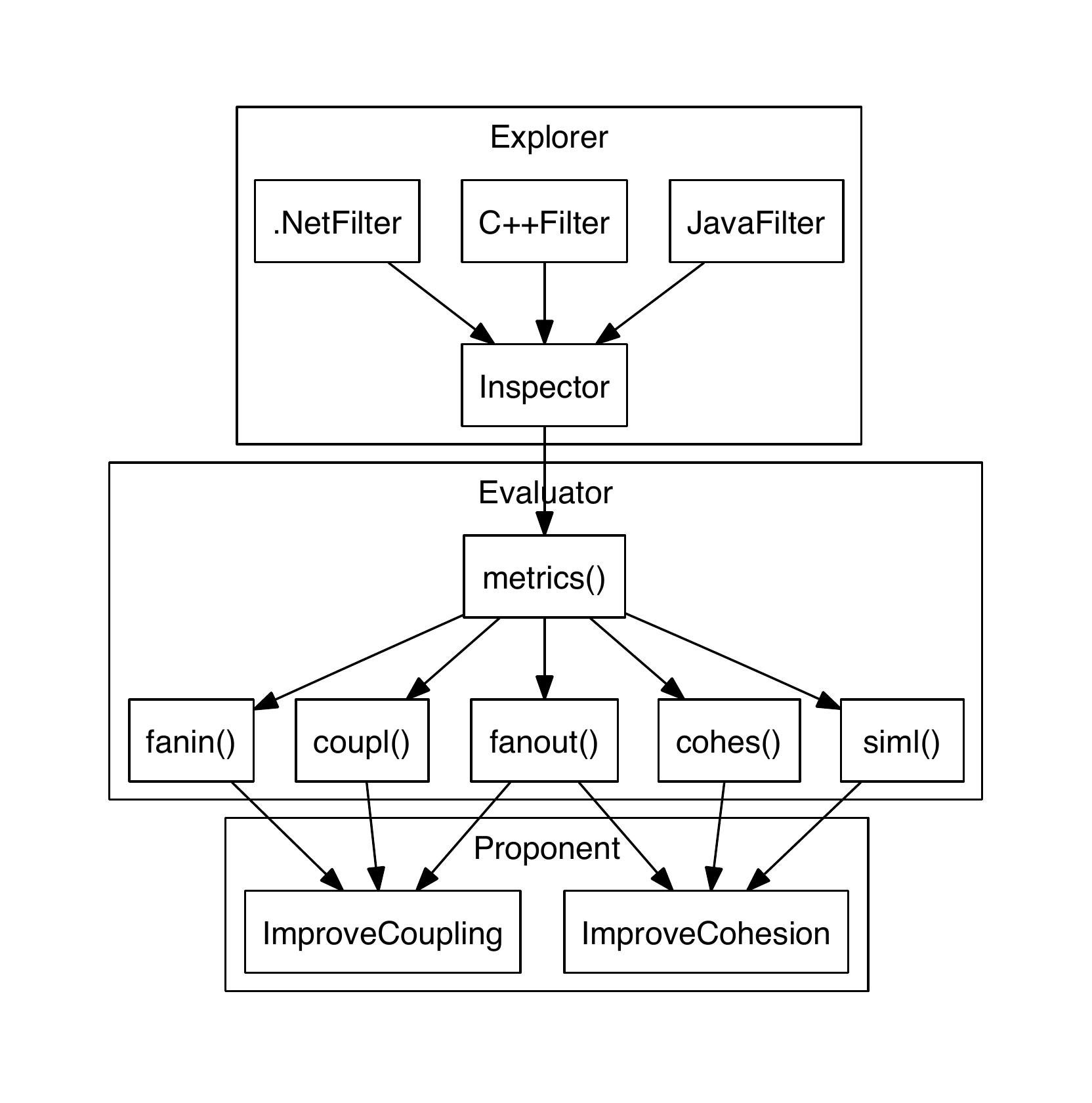}
    \caption{Main components of the tool for the proposed approach}
    \label{graphTool}
  \end{center}
\end{figure}

Figure~\ref{graphTool} provides an overview of the architecture for a
tool computing the above metrics. Firstly, the system under analysis
will be explored and for this several filters are needed to
recognise the characteristics of the system.
Once exploration has been performed, metrics will be computed,
according to extracted characteristics and the definitions of metrics.
Finally, metrics are used together to suggest some move method
refactoring.  Each refactoring could improve a subset of metrics only,
hence a desired characteristic, such as e.g.\ cohesion or
coupling. The developer could select the characteristic to improve or
ask for refactoring suggestions that improve all of them. In the
latter case, a smaller subset of suggestions would be given, i.e.\ the
intersection of refactoring suggestions.

The following subsections provide details of both the exploration of
the system and metric computation on a GPU, whereas the logic for
selecting a refactoring has been described in Section~\ref{approach}.

\subsection{Exploring the system under analysis}
%Extracting Data}
Metrics described in Section~\ref{metrics} are mainly based on the
identification of attributes, methods and classes of a software
system, as well as the number of method calls and attribute accesses
from each method of every class.
In order to compute the above metrics, firstly such data will have to
be gathered.
Several approaches can be followed for pursuing this
\emph{exploration} task, e.g.\  computational
reflection, as in~\cite{PacoSac06}, or supporting libraries.
For our experiments, we have used the Bytecode Engineering Library
(BCEL) for gathering data on the bytecode of a Java software system,
hence greatly simplifying the exploration
phase~\cite{dahm98}\footnote{http://commons.apache.org/bcel/}.
BCEL can be considered as representing the \textsf{JavaFilter} block shown in
Figure~\ref{graphTool}. Other filters can be built as desired for
different languages or executable code. 
%%% Paradyn is a powerful tool for instrumenting C++

We have then built an \textsf{Inspector} that gathers all the classes
names of a software system, and for each class analyses the method
bodies.  Firstly, classes names, attributes and methods names have
been found and stored into a list. For attributes and methods the name
of the class they belong to has been stored. Moreover, methods names
have been stored together with their input parameters, in order to
properly consider overloaded methods.

Every method body has been analysed and method calls and attribute
accesses have been stored into another list. Such a list can be
navigated when accessing a method (on the previous list) in order to
have all the called methods and accessed attributes.  This is the main
part of \textsf{Inspector}, finding dependencies between classes and
between methods.

\subsection{Computing metrics on a GPU}
% \label{gpgpu}

% A GPGPU consists of several MIMD (multiple instruction multiple data)
% multiprocessors each containing a set of SIMD (single instruction
% single data) processors. Each MIMD multiprocessor is equipped with a
% shared memory that can be accessed from each of its SIMD processors,
% and a global memory common to all multiprocessors.
%
In CUDA programming model, an application consists of a \emph{host}
program that executes on the CPU and other parallel \emph{kernel}
programs executing on the GPU~\cite{nvidia}.
%et: dove e' pubblicato questo ultimo?
%
A \emph{kernel} program is executed by a set of parallel threads.
The \emph{host} program can dynamically allocate device \emph{global}
memory to the GPU and copy data to (and from) such a memory from (and
to) the memory on the CPU.  Moreover, the \emph{host} program can
dynamically set the number of threads that run on a \emph{kernel}
program.
Threads are organised in blocks, and each block has its own
\emph{shared} memory, which can be accessed only by each thread on the
same block.

For maximum performances, threads on a GPU should ideally be given a
task that can run unconstrained, i.e.\ without having to synchronise
with others~\cite{gems}.  Moreover, it is paramount that interactions
between CPU and GPU are minimised, this avoids communication
bottlenecks and delays due to data transfers.

Following such general guidelines, we have developed a program,
realising block \textsf{Evaluator} in Figure~\ref{graphTool}, that
takes as input the list of classes, their methods and attributes, of
the software system to be analysed. For handling the minimum possible
amount of data, classes, methods and attributes have been represented
by a numerical id.
We have used arrays for holding several ids, since they can be easily
and efficiently passed to the GPU, i.e.\ the \emph{host} program
allocates memory and transfers data to such a memory, which CUDA
\emph{kernel} program can use.
For the classes, array \texttt{cls} holds the ids of each class, as
well as the corresponding ids of methods and attributes.
For methods, an array \texttt{dpnd} holds for each method its
\emph{dependencies} in terms of the ids of methods invoked and
attributes accessed.
% %
% For concerns, an array \texttt{grps} holds for each concern the
% classes belonging to it. 

By calling the standard \texttt{cudaMemcpy()} function,
arrays \texttt{cls} and \texttt{dpnd} are passed to the GPU memory,
hence they become
 available 
to our provided \emph{kernel} global function
\texttt{metrics()}.  This calls other \emph{kernel} device functions,
 each  computing one of the different metrics 
used.  Among the said arrays, only appropriate ones are given to the function specialised
for computing one of the metrics.  The values of arrays are read by each
thread, however threads need not write any value on the arrays, hence
no synchronisation has been used for accesses.

For the function computing the similarity metric, \texttt{siml()},
each available thread is given a range of method ids, representing a
subset of all the available methods to be analysed.  For the given
range of ids, a thread executing inside our function \texttt{siml()}
computes all the similarity values between one method and all the
other methods, by reading values from array \texttt{dpnd}, providing
data representing dependencies.
The selection of the range of methods to be given to a thread is
easily determined by the maximum number of methods available and
\texttt{ThreadId}, available in CUDA programs, indicating the current
working thread.

Each thread stores results into its own local array, 
% Note that having each thread store its own result values 
i.e.\ separately from other threads, hence minimising the need of 
%. This allows  threads to run unconstrained of 
synchronisation.  
%Of course, 
Given the
large amount  method pairs (see the above analysis~\ref{scales},
and the
% examples of
 the analysed systems~\ref{caseStudies}), only
meaningful values are stored, i.e.\ only similarity values that are
greater than zero (otherwise we risk filling up all 
available memory).
Once a thread has finished executing, it will have computed and stored
a given amount of results, which likely differs in number from that of
other threads.  This is because each thread will find a different
% an a priori unknown 
number of zeros as a result. The meaningful values will have to be
stored on a globally accessed array, so that other functions on the
device can use them.  For this, each thread reserves an amount
of locations to store its computed values.  Reservation has been
performed by updating a global variable, shared by threads, hence by
using the \texttt{atomicAdd()} function. This is the only moment for
threads to synchronise with each other.

For metric LCOM, our function \texttt{cohes()} is given a range of
ids for classes, hence each thread computes values of LCOM
%for each class
for a subset of classes by reading values from arrays \texttt{cls} and
\texttt{dpnd}.
Like the previous function, \texttt{siml()}, each thread stores the
non zero value of LCOM for a class locally, before processing the
following class.  Once all classes have been processed, relevant
results are transferred to the global array by resorting to a
synchronous update of a shared variable.
Similarly, functions have been implemented for each other
metric, i.e.\ fan-in, fan-out, and CBO.

For assessing the benefits of methods moves refactoring, candidate
classes to be changed, both as an origin or a destination class for a
moved method, will have to be examined and their relevant changing
values, i.e.\ LCOM and CBO computed again.  This is performed by
changing the representation of the methods belonging to a class, and
then by repeating execution within the said functions. 
%, e.g.\  \texttt{cohes()}.

\begin{table*}
  \caption{Analysed software systems}
  \label{TabSystems}
  \centering
  \begin{tabular}{ r l | r  r | r   r  r  r |  r r r | r}
    \hline
\multicolumn{2}{c|}{Systems} & LOC & NCNB & Classes & Attributes&
Methods &  \#values & \multicolumn{3}{c|}{execution time} & \#move\\ 
    &&&&& & & & CPU&Tesla&ratio & methods\\ 
    \hline
1 & JUnit                  &30K&22K&    231&   265& 1,200& 1.2M&0.10&0.07&1.53&2\\
2 & JHotDraw           &72K&28K&   600& 1,151&4,814& 17.4M&1.61&0.30&5.43&26\\
3 & JavaStyle           &69K&26K&   600& 1,423& 6,816&31.4M&5.55&0.24&22.74&54\\
4 & Hammurapi        &80K &30K& 986& 2,595& 7,705&44.9M& 9.73&0.24&40.54&167\\
5 & Dependometer   &75K &32K & 907& 2,932& 7,858&45.1M& 7.43&0.41&18.03&632\\
6 & MapperXML        &42K&16K& 1,146& 2,726& 8,074&51.1M&10.26&0.31&33.31&687\\
7 & JEdit                  &183K&77K& 1,267& 3,804& 9,629&70.8M&13.31&0.56&23.93&309\\ 
8 & Commons-math&276K&115K&1,930&4,196&13,676&146.3M&30.54&1.10&27.86&367\\ 
9 &  Weka               &529K&206K&2,138& 9,194&22,028&336.8M&80.79&1.63&46.62&435\\
10 & JRefactory &302K&120K&2,775&6,053&23,639&410.6M&66.10&1.65&40.01&864\\
11 & Derby          &1174K&403K&3,191&13,900&44,394&1,268.8M&256.51&4.68&54.81&887\\
12 &Libomv    &195K& 73K&7,134&14,211&43,593&1,572.2M&218.45&4.80&45.51&438\\
%  Cdk     &       &        &4,728&13,227&44,696&1,632.9M&249.70&16.50&15.14&438\\
13 & ProjectLibre  &465K&174K&6,399&28,444&69,751&3,325.4M&569.75&12.94&44.04&2,358\\
%  Hibernate       & & &12,575&34,524&104,247&9,366.6M&\\
    \hline
  \end{tabular}
\end{table*}

\section{Evaluation}
\label{caseStudies}

% In our experiments w
Table~\ref{TabSystems} provides values of several structure
characteristics 
%of the analysed systems.
and 
%We have 
the computed the metrics 
%given in Section~\ref{metrics} 
for several software systems, ranging from small
to medium size systems.
%
% \footnote{Given the size of the
%   analysed systems, we cannot provide the values of object-oriented
%   metrics for each class.}.
%
The size of analysed systems is given as the number of lines of code
(LOC) as well as the number of lines no comment no blank (NCNB). The
number of total classes is shown along with the number of attributes
and methods for each class.
Column \#values shows the amount of values for the several metrics
that have to be computed, according to the analysis given in
Section~\ref{approach}.
It can be seen that such values grow to the order of thousands of
millions.

\begin{figure}
  \begin{center}
    \includegraphics[scale=0.34]{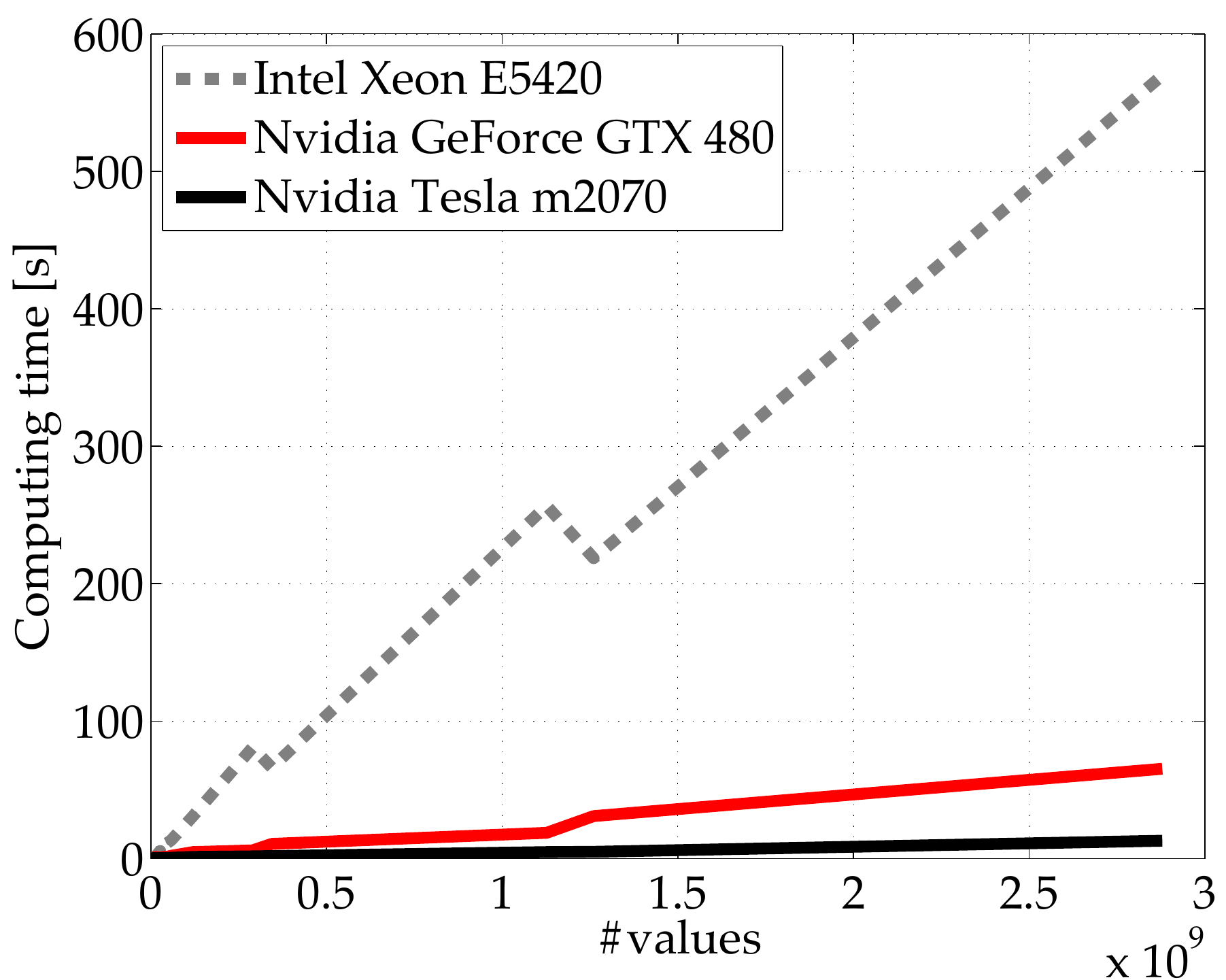}  %0.38 width=\columnwidth
    \caption{Measured computing times for analysed systems}
    \label{plotTime}
  \end{center}
\end{figure}

\begin{figure}
  \begin{center}
    \includegraphics[scale=0.34]{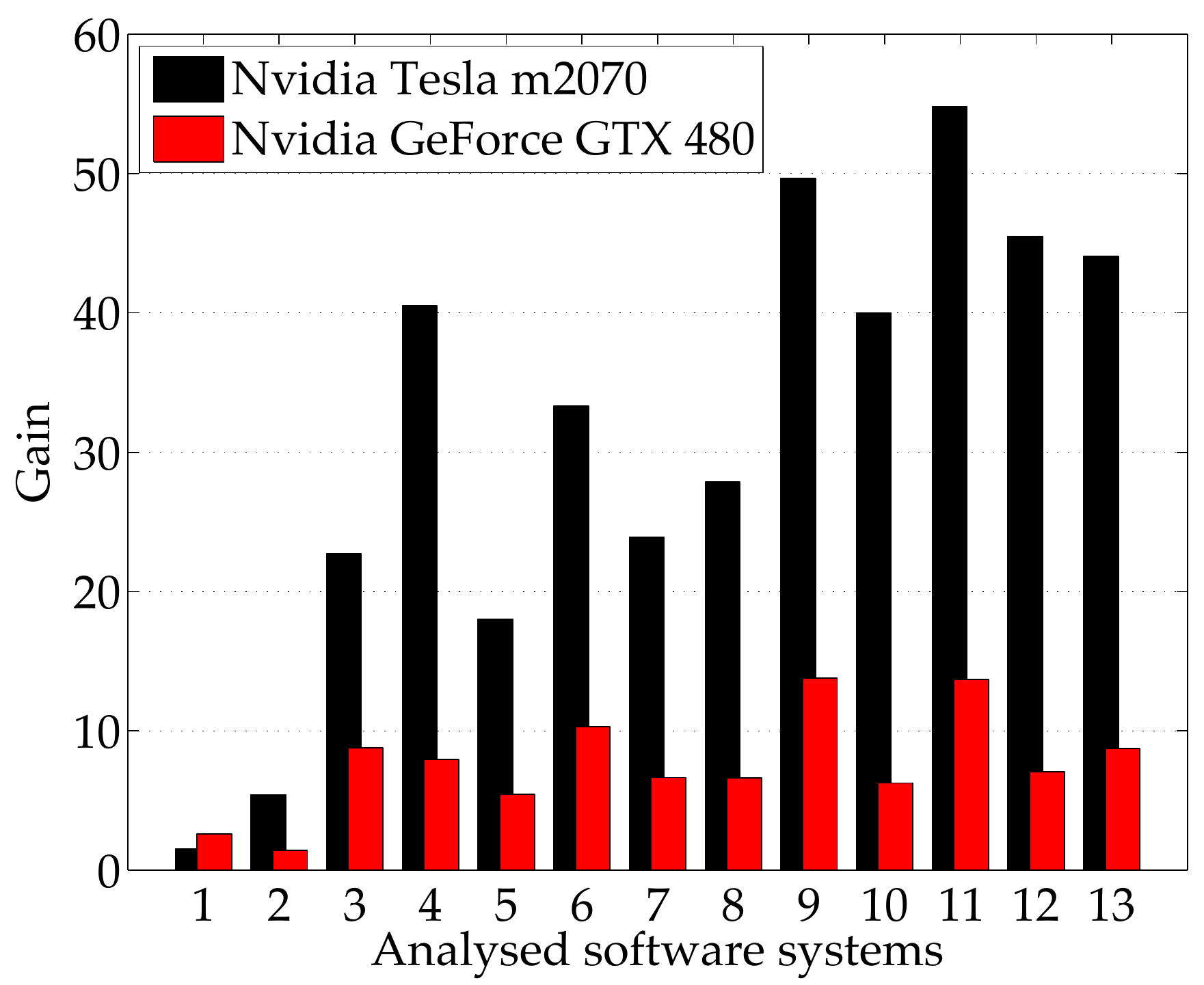} 
    \caption{Gain ratio between GPUs and CPU computing times}
    \label{barsGain}
  \end{center}
\end{figure}

The three columns \emph{execution time} of Table~\ref{TabSystems} show
measured wall times when computing metrics for improving cohesion
possibly for all classes, i.e.\ values of similarity, LCOM, and again LCOM
for classes involved into move method opportunities, as in block
\textsf{ImproveCohesion} within \textsf{Proponent} in
Figure~\ref{graphTool}.  The wall times refer to a single CPU and on a
Tesla GPU, with 448 cores, and their ratio. As we can see also from
Figure~\ref{plotTime}, computing times go from minutes to seconds for
larger systems passing from a CPU to a GPU (from 11 min to 13
secs for the largest system). The plot comprises
computing times needed with both a Tesla and a GeForce GPUs.  Note
that when comparing the gain CPU over Tesla and the gain CPU over
GeForce, the latter is smaller, however still very significant (and
with a fraction of the price for the hardware needed).
The gain in performance shows important improvements ranging from a
$1.5$ gain for a small system to a $54.8$ gain for one of the largest
systems analysed (see Figure~\ref{barsGain}).
The gain that we have obtained is in good agreement with previous
assessments of other programs, when we compare an appropriate solution
using GPU resources with an optimised solution on a
CPU~\cite{VictorLee2010}.

Finally, the last column in Table~\ref{TabSystems} shows the number of
move methods refactoring that have been automatically suggested by our
tool for each software system under analysis when considering move
methods refactoring that optimise LCOM values only.

From the number of suggestions, reaching more than 2 thousand for the
largest system, it is possible to conclude that the high-quality
systems analysed can be actually further improved thanks to the
 selection of move method opportunities proposed by
our approach.
By enhancing modularity, we make systems more prone to incorporate
other functional and non-functional requirements. E.g.\ less
interactions between components could facilitate consistent runtime
updates~\cite{BannoMPT10a}.

%% \section{Conclusions}
\section{Conclusions}
\label{conclusions}

This paper has proposed several criteria for suggesting move methods
refactoring opportunities that aim at improving several values of
modularity metrics for a software system.
It has been shown that for obtaining a proper view on the effect of changes for
a large software system the number of values to be calculated grows
very quickly. This can become overwhelming for a single CPU, hence a GPU
can be purposely employed.
It is also important to notice that refactoring opportunities can be
more difficult to assess for the unassisted developer when having to
reason on a large system. Hence, a tool that readily checks thousands
of millions of values is of great help.
Experiments have shown that existing large systems can be further
\appendices
\section*{Acknowledgment}
This work has been supported by project Infinity Web Edition funded within POR FESR Sicilia 2007-2013 framework, and project PRISMA PON04a2 A/F funded by the Italian Ministry of University. 
\\ ~\\
This paper has been published in the final and reviewed version
on {\bf 7th International Conference on Complex, Intelligent, and Software Intensive Systems (CISIS), pp.
529-534, 2013} \cite{rr}.

\bibliographystyle{IEEEtran}
\bibliography{napoliOA13ICLS1}

\end{document}